# Elastic Wave Propagation and Bandgaps in Finitely Stretched Soft Lattice Material

Shiheng Zhao, [a, †] Tao Feng, [a, †] Han Zhang, [b] Yang Gao, [a] and Zheng Chang [a,*]

[a] College of Science, China Agricultural University, Beijing 100083, People's Republic of China

[b] Key Laboratory of Noise and Vibration, Institute of Acoustics, Chinese Academy of Sciences, Beijing 100190, People's Republic of China

## Abstract

In this study, the in-plane Bloch wave propagation and bandgaps in a finitely stretched square lattice were investigated numerically and theoretically. To be specific, the elastic band diagram was calculated for an infinite periodic structure with a cruciform hyperelastic unit cell under uniaxial or biaxial tension. In addition, an elastodynamic "tight binding" model was proposed to investigate the formation and evolution of the band structure. The elastic waves were found to propagate largely under "easy" modes in the pre-stretched soft lattice, and finite stretch tuned the symmetry of the band structure, but also "purify" the propagation modes. Moreover, the uniaxial stretch exhibits the opposite impacts on the two "easy" modes. The effect of the biaxial stretch was equated with the superposition of the uniaxial stretches in the tessellation directions. The mentioned effects on the band structure could be attributed to the competition between the effective shear moduli and lengths for different beam components. Next, the finite stretch could tune the directional bandgap of the soft lattice, and the broadest elastic wave bandgaps could be anticipated in an equi-biaxial stretch. In this study, an avenue was opened to design and implement elastic wave control devices with weight efficiency and tunability. Furthermore, the differences between the physical system and the corresponding simplified theoretical model (e.g., the theoretically predicted flat bands) did not exist in the numerical calculations.

[†] These authors contributed equally to the study.
[*] Corresponding author. E-mail address: changzh@cau.edu.cn (Z. Chang).

# 1. Introduction

Lattice materials (Phani and Hussein, 2017) are defined as the spatially periodic networks of structural elements (e.g., beams, bars and shells). With a highly ordered construction at different scales, such materials outperform conventional (dis-order) ones in mechanical (Schaedler et al., 2011; Zhang et al., 2014; Zhang et al., 2015; Casalotti et al., 2020), thermal (Wei et al., 2016; Yun et al., 2020) and other properties (Huntington et al., 2014; Xia et al., 2019; Montgomery et al., 2020). These optimized structural and functional materials appear extensively in nature (Gibson et al., 2006; Guo et al., 2018; Fernandes et al., 2021), and meanwhile are artificially manufactured and increasingly used as the weight-efficient structural components in diverse applications (Chen et al., 2017b; Chen et al., 2021).

Over the past two decades, lattice materials have aroused considerable attention for their unique elastodynamic properties that originate from the periodic arrangement and variety of microstructures. These materials, as a class of phononic crystals or metamaterials, have been commonly employed to tailor the propagation of mechanical waves. As inspired by this topic, the novel wave phenomenon of lattice materials (Phani et al., 2006; Trainiti et al., 2016; Phani and Hussein, 2017; Bordiga et al., 2019; Bordiga et al., 2021) and the effects of both material and geometry (Huang et al., 2018; Bordiga et al., 2021) have been investigated. Thus far, the dynamics of lattice materials may be considered a mature research field, since the governing equations and the solution techniques are generally known. However, many exciting characteristics remain unclear, as admitted by the diversity of material and structure combination and driven by various technological requirements.

For instance, if the solid phase exhibited by the lattice material is soft and highly deformable, more complex but fruitful wave behaviors can be anticipated under the use of the external mechanical loading. Accordingly, the lattice turns out to be a sophisticated wave modulation and sensing system, which suggests significant applications (e.g., the non-destructive testing and the health monitoring for flexible and morphable structures). The mechanical loading can tense the network, and it acts as a robust mechanism for tunability in the preparation and use of the lattice material. As demonstrated by a spider orb-web, i.e., an ingenuity of nature, it can form a mechanically stable construction with taut threads (Wirth and Barth, 1992). Moreover, it transmits vibration signals, so it acts as a communication channel during the prey capture and the courtship (Landolfa and Barth, 1996; Mortimer et al., 2016). Tensions in an orb-web are not evenly distributed, and they can be controlled by a spider (Wirth and Barth, 1992) when the web is being weaved and used (Spider jerks the radial threads to locate the prey.).

Though the dynamics of soft phononic crystals and metamaterials have been extensively studied, most of the relevant works (Bertoldi and Boyce, 2008; Bilal et al., 2020; Muhammad et al., 2020; Dal Poggetto and Serpa, 2021) concentrated on materials exhibiting high relative densities or with over one material phase. As opposed to the mentioned, the low-density and single-phase lattice materials have been rarely studied, which is usually more lightweight, flexible and easy to fabricate. Though some works revealed the biological functions and wave modulation mechanisms (Frohlich and Buskirk, 1982; Mortimer et al., 2016) of natural lattices, there has been rare source allowing for reasonably overviewing the design criteria of bio-inspired lattice materials. To be specific, how the mechanical loading simultaneously affects the effective (or instantaneous) material properties and geometry configuration of the soft lattice and how the mentioned two factors impact wave propagation remain unclear. In the existing work (Zhao and Chang, 2021) on the dynamics of a single soft beam, the authors suggested that the wave propagation characteristics (e.g., wave phase velocities) could be highly determined by the pre-stress. Moreover, longitudinal and transverse waves manifested different responses to external loading. For this reason, the dynamics of the soft lattice composed of such soft beams is worth investigating.

To clarify the mentioned problems, this study should determine the effects of geometric and material properties induced by the finite deformation on the linear in-plane elastic wave response of a 2D soft (hyperelastic) lattice structure. In accordance with the Small-on-Large theory (Ogden, 2007), the numerical simulations were performed to examine the formation and evolution of the band structure. Moreover, an elastodynamic "tight binding" model was proposed to provide valuable estimates and elucidate the underlying mechanisms.

The rest of this study is organized below. In Section 2, the model description and several theoretical backgrounds are presented (e.g., the Small-on-Large and Bloch theories). In Section 3, the numerical and theoretical methods of examining the band structure of the soft lattice are illustrated. In Section 4, the band structure, the effect of pre-deformation, together with the evolution of bandgaps are illustrated. Lastly, the brief concluding remarks are drawn in Section 5.

## 2. Model description and theoretical background

### 2.1. Model description

An infinite 2D square lattice was considered, with its initial configuration $\Omega$ presented in Fig. 1(a). The unit cell was a cruciform structure rigidly connected by four flexural beams, with the related tessellation directions (lattice basis vectors) parallel to the beams' axes. The material of the respective beam was homogeneous, isotropic and soft, so it exhibits a hyperelastic constitution. In the plane-strain assumption, the finite deformation and the incremental wave motion were constrained in the $X_1X_2$ plane. The lattice's macroscopic deformation was considered under the uniaxial or biaxial tensile strains in the tessellation directions for simplicity. Accordingly, the orthogonal beams hardly affected each other during the deformation, and any possible instability induced by the compression was avoided.

After the finite deformation, the soft lattice expanded to the current configuration $\Omega'$ (Fig. 1(b)), and the corresponding unit cell emerged a cruciform with diverse arm lengths (Fig. 1(c)). Except for the joint of the beams, the most area in the unit cell exhibited the state of uniaxial tension. In this fashion, the finite deformation could be expressed by the principal stretches ($\lambda_1$ and $\lambda_2$). Furthermore, the geometry of the current configuration could be characterized by the lengths ($l_{x_1}$ and $l_{x_2}$) and widths ($h_{x_1}$ and $h_{x_2}$) of the beams.

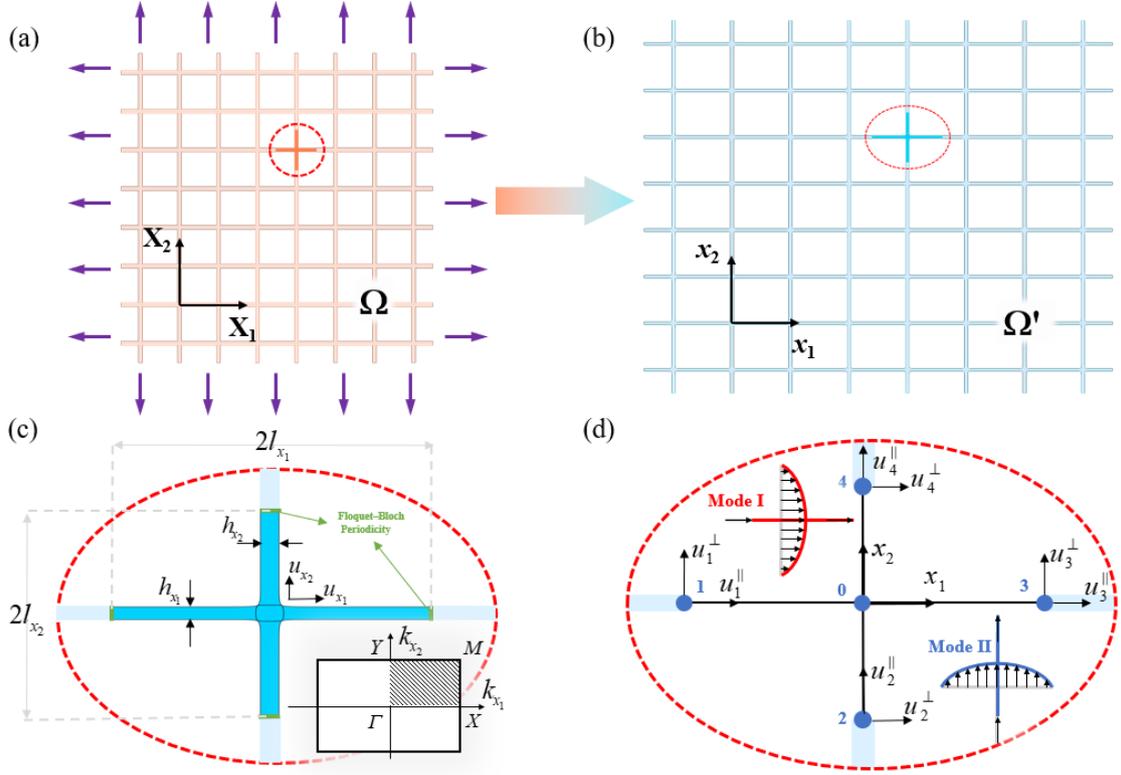

Fig. 1. Schematic diagrams of the 2D soft lattice material and its unit cell. (a) and (b) represent the initial ($\Omega$) and current ($\Omega'$) configurations before and after the finite deformation, respectively. (c) and (d) represent the unit cells of the current configuration adopted for the numerical and theoretical analyses, respectively. The first Irreducible Brillouin Zone (IBZ) of the unit cell applied in the Bloch wave analysis is presented in the inset of (c). The two "easy" wave modes are illustrated in the insets of (d).

## 2.2. Theoretical background

To analyze the propagation of the Bloch wave in the soft lattice, some preliminary (e.g., the descriptions of finite deformation, incremental wave motion, and the Bloch boundary condition) are presented in the following, which are essential for both numerical and theoretical analyses.

### 2.2.1. Finite deformation

For a hyperelastic solid with the constitutive behavior characterized by the strain energy function $W$, the finite deformation $U_i$ mapped the initial configuration ($\Omega$) into the current one ($\Omega'$), i.e., $U_i = x_i(X_j)$, where $x_i$ and $X_j$ respectively denote the Cartesian components in the current and initial configurations. In the absence of the body force, the finite deformation satisfied the equilibrium equation, i.e.,

$$\mathcal{S}_{ij,i} = 0, \tag{1}$$

where $\mathcal{S}_{ij} = \partial W/\partial F_{ji}$ denotes the nominal (or first *Piola-Kirchhoff*) stress tensor; $F_{ij} = \partial x_i/\partial X_j$ is the deformation gradient (unless otherwise specified, the repeated lower index complied with the Einstein summation convention). From the relationship between stress and deformation gradient, the instantaneous elastic tensor (or tangent modulus) of the material was written as

$$C_{ijkl} = \frac{\partial \mathcal{S}_{ij}}{\partial F_{lk}} = \frac{\partial W}{\partial F_{ij} \partial F_{lk}}. \tag{2}$$

Thus, Eq. (1) can be expressed for $U_i$ as

$$(C_{ijkl} U_{l,k})_{,i} = 0. \tag{3}$$

Since $C_{ijkl}$ was non-linearly determined by $F_{ij}$, Eq. (3) expresses a set of non-linear partial differential equations of the second order for $U_i$. Specific to a particular $U_i$ (or equivalently $F_{ij}$), the uniquely determined $C_{ijkl}$ was termed as the instantaneous elastic tensor expressed in the initial configuration.

$C_{ijkl}$ and $U_i$ were utilized to analyze the incremental wave motion.

*2.2.2. Incremental wave motion*

To examine the incremental (linear) wave motion $u_i$ superimposed onto the finite deformation $U_i$, the stress generated by the waves was neglect, the finite prestress was equated with the instantaneous material parameters (Eq. (2) and Eq. (5)), and the waves were considered to propagate in a stress-free effective medium. Thus, the

governing equation of wave motion in the current configuration is expressed as (Ogden, 2007)

$$(\tilde{C}_{ijkl}u_{l,k})_{,i} + \tilde{\rho}\omega^2 u_j = 0, \tag{4}$$

where $\omega$ denoted the angular frequency. $\tilde{C}_{ijkl}$ and $\tilde{\rho}$ were determined by pushing forward on $\tilde{C}_{ijkl}$ and $\rho$ (expressed in the current configuration), respectively, (Ogden, 2007)

$$\tilde{C}_{IjKl} = J^{-1}F_{Ii}F_{Kk}C_{ijkl}, \tilde{\rho} = J^{-1}\rho, \tag{5}$$

where $J = \det(F_{ij})$ denotes the volumetric ratio. As opposed to Eq. (3), Eq. (4) expresses a set of linear partial differential equations of the second order for $u_i$.

### 2.2.3. Bloch boundary condition

To determine the spatial periodicity of the current configuration, a set of lattice vectors $\vec{a}_s$ ($s = x_1, x_2$) were defined, so an arbitrary unit cell could be indexed with the vector $r_s = n_{\underline{s}}|a_{\underline{s}}|$, with $n_s$ as an integer (the underlining represents that the not applied Einstein summation convention). Accordingly, the position of any point $x_s^0$ in a unit cell was expressed as $x_s = x_s^0 + r_s$. By complying with the Bloch theorem, the Bloch periodicity for arbitrary space function $\psi$ was defined as:

$$\psi(x_s^0 + r_s) = \psi(x_s^0)e^{ik_s r_s}, \tag{6}$$

where $\mathbf{i}$ denotes $\sqrt{-1}$; $k_s$ is the wave vector in the reciprocal lattice (Fig. 1(c)).

Eq. (4) and (6) usually formed a definite solution problem of wave propagation in a periodic soft lattice, which underpinned the following numerical and theoretical analyses.

## 3. Methods for Bloch wave analysis

In the present section, the numerical method is presented to determine the band structure of the soft lattice. Moreover, a theoretical model was proposed to analyze the effect, together with the underlying mechanism, of the finite deformation on the Bloch wave behavior.

### 3.1. Numerical (Finite Element) analysis

The numerical simulations were conducted with the finite-element (FE)-method-based software COMSOL Multiphysics. A two-step model (Chang et al., 2015; Chen et al., 2017a) was adopted to analyze the finite deformation and the incremental wave motion, respectively.

First, the finite deformation of the hyperelastic material, complying with Eq. (1), was calculated with the module of solid mechanics. For the soft lattice expressed in Section 2.1, the analysis was conducted based on a unit cell in the initial configuration $\Omega$ (Fig. 1(a)). Specific to the uniaxial stretch in the $x_1$ (or $x_2$)-direction, the prescribed displacement along the beam axis was introduced to the unit cell's right (or upper) end. At the opposite end, a roller condition was imposed to constrain the axial displacement, while the lateral deformation was not affected. Moreover, the midpoint of the roller was fixed to prevent the rigid body translation, and the rest boundaries were free of constraint. Under the biaxial stretch, with the mentioned settings unchanged, the upper and lower boundaries were replaced with the prescribed displacements exhibiting the identical magnitude but in opposite directions. Thus, the deformed configuration of the unit cell (Fig. 1(c)), together with the corresponding deformation field, could be determined through the nonlinear quasi-static analysis.

Second, the band structure of the lattice was determined with the module of weak-form PDE. With the obtained deformed geometry (Fig. 1(c)) and the deformation gradient imported, Bloch periodicity (Eq. (6)) was imposed at the four ends of the unit cell. Based on an eigenvalue analysis of Eq. (4), the band structure, i.e., the dispersion relation of elastic waves, was explored along the edges ($\Gamma - X - M - Y - \Gamma$) of the first Irreducible Brillouin Zone (IBZ, Fig. 1(c)).

### 3.2. Theoretical model: an elastodynamic "tight binding" model

Despite the numerical simulation, a theoretical model was also built to characterize the Bloch wave propagation, and more importantly, to qualitatively indicate the effect of finite deformation on the band structure. The irregular geometry and the inhomogeneous deformation in the unit cell, mainly distributed close to the junction, were neglected in this study. Thus, the unit cell was simplified as a nodal connection of four beams under the ideal uniaxially stretch. As illustrated in Fig. 1(d), the internal node was numbered as $0$, and the peripheral nodes were numbered as $1-4$, respectively. The indexes of the four beams were identical to the corresponding peripheral nodes. Thus, $\{u_i^{\parallel}(s), u_i^{\perp}(s)\}, i=1,2,3,4$ was adopted to represent the axial and transverse displacements (relative to the beam direction) in the $i$-th beam, where $s$ denotes the local axial coordinate ($s=x_1$ (or $x_2$) for $i=1, 3$ (or $2, 4$)) with the origin at the central node (Fig. 1(d)).

### 3.2.1. Equations of motion in a single beam

For a single beam under the axial stretch, there existed two wave modes, i.e., the longitudinal (Primary, or P- for short) and transverse (Secondary, or S-) modes, which were respectively governed by the one-dimensional time-harmonic wave equations (Piccolroaz et al., 2017; Bordiga et al., 2019; Zhao and Chang, 2021)

$$\tilde{E}u_i^{\parallel\prime\prime}(s)+\tilde{\rho}\omega^2 u_i^{\parallel}(s)=0, \tag{7}$$

$$B\tilde{E}u_i^{\perp\prime\prime\prime\prime}(s)+\left(B\tilde{\rho}\omega^2-\tilde{G}\right)u_i^{\perp\prime\prime}(s)-\tilde{\rho}\omega^2 u_i^{\perp}(s)=0, \tag{8}$$

where the prime represents the differentiation with respect to $s$. $B=I/A$ expresses the moment of inertia divided by the cross-sectional area; $\tilde{E}$ and $\tilde{G}$ denote the effective Young's and shear moduli, respectively.

Specific to a 2D problem, $\tilde{E}$ and $\tilde{G}$ were expressed with the components of the push forwarded effective elastic tensor as (Zhao and Chang, 2021):

$$\tilde{E} = \tilde{C}_{1111} - \frac{\tilde{C}_{1122}\tilde{C}_{2211}}{\tilde{C}_{2222}},$$

$$\tilde{G} = \tilde{C}_{1212} - \frac{\tilde{C}_{1221}\tilde{C}_{2112}}{\tilde{C}_{2121}}. \tag{9}$$

A similar but more complex expression was obtained for a 3D problem. It is noteworthy that Eq. (9) was derived for the ideal uniaxial stretch on a single beam. However, such a condition could not be fully satisfied by the beam components in the lattice material attributed to their cross-shaped spatial connection as well as the resulting inhomogeneous deformation around the junction. Thus, in the theoretical model, $\tilde{E}$ and $\tilde{G}$ were calculated using the deformation gradient of a beam component obtained from the numerical simulation (Supplementary Material). On that basis, the obtained effective parameters were suggested to be larger than those calculated for a single beam, and the lattice material's wave propagation was more accurately predicted.

### *3.2.2. Boundary conditions*

Twenty-four boundary conditions were required to establish a proper combination of equations of motion for each beam, which included the following:

a). Nine kinematic conditions at the internal node, representing the compatibility of displacement, i.e.,

$$u_1^p(0) = u_2^q(0) = u_3^p(0) = u_4^q(0), (p,q = \perp, \|, p \neq q), \tag{10}$$

and rotation (with rigid joint assumption), i.e.,

$$u_1^{\perp\prime}(0) = u_2^{\perp\prime}(0) = u_3^{\perp\prime}(0) = u_4^{\perp\prime}(0); \tag{11}$$

b). Three equilibrium conditions at the internal node, i.e.,

$$N_i(0) - N_{i+2}(0) + V_j(0) - V_{j+2}(0) = 0, (i, j = 1, 2, i \neq j) \tag{12}$$

$$M_1(0) - M_3(0) + M_2(0) - M_4(0) = 0, \tag{13}$$

where

$$\begin{aligned} V_i(s) &= -\tilde{E}_s A_s B_s u_i^{\perp\prime\prime\prime}(s) + \left(\tilde{G}_s - \tilde{\rho}_s B_s \omega^2\right) A_s u_i^{\perp\prime}(s), \\ N_i(s) &= \tilde{E}_s A_s u_i^{\|\prime}(s), \\ M_i(s) &= \tilde{E}_s I_s u_i^{\perp\prime\prime}(s), \end{aligned} \tag{14}$$

denote the shear force, the axial force and the bending moment in the $i$-th beam, respectively;

c). Twelve Bloch boundary conditions at the peripheral nodes, representing the periodicity of the displacement, rotation, bending moment and shear force. In the local coordinate system, the Bloch conditions (Eq. (6)) were expressed as:

$$\Psi_{i+2}\left(x_s^0 + r_s\right) = \Psi_i\left(x_s^0\right) e^{ik_s r_s}. \tag{15}$$

To be specific, we take $\Psi_i = u_i^\perp, u_i^{\perp\prime}, u_i^\parallel, u_i^{\parallel\prime}, M_i, V_i$, $x_s^0 = -l_s$ and $r_s = 2l_s$ in Eq. (15). ($s = x_1$ (or $x_2$) for $i = 1, 3$ (or 2, 4))

### 3.2.3. Assembly of equations of motion

With eight equations of motion (Eq. (7), (8)) and twenty-four boundary conditions, an eigenvalue problem was formed with a $24 \times 24$ characteristic matrix (Bordiga et al., 2019). In this study, more accurate results could be achieved by performing the numerical treatment. The theoretical model was mainly used to reveal the underlying mechanism. However, such a huge algebraic system could be too complicated to yield an analytical solution concise enough for qualitative analyses. As a trade-off, instead of solving the characteristic determinant directly, the algebraic system was first degenerated into two smaller ($12 \times 12$) systems by confining (or "binding") the possible deformation of the soft lattice into two "easy" (easiest to be excited, or fundamental) modes. Consistent with the tight-binding model in solid physics, this degenerate model was termed as an elastodynamic "tight-binding" model. Given the mode shapes illustrated in the insets of Fig. 1(d), the two beams with the identical orientation in the unit cell vibrated in a P- or S- mode, whereas the other two beams presented the counterpart mode. These mentioned modes led to the identical displacement directions of the respective material point in the unit cell, thereby exhibiting maximum structural compliance.

The elastodynamic "tight-binding" model required the rigid joint constraint at the internal node to be relaxed to construct the degenerated problem. Accordingly, the compatibility condition of rotation (Eq. (11)) was replaced with:

$$u_1^{\perp\prime}(0) = u_3^{\perp\prime}(0), \tag{16}$$

$$u_2^{\perp\prime}(0) = u_4^{\perp\prime}(0). \tag{17}$$

In correspondence, the equilibrium of bending moment (Eq. (13)) should be replaced with:

$$M_1(0) = M_3(0), \tag{18}$$

$$M_2(0) = M_4(0). \tag{19}$$

Thus, under Mode I (Fig. 1(d)), the corresponding equations of motion were Eq. (7) for $i = 1, 3$, and Eq. (8) for $i = 2, 4$. Moreover, the selected boundary conditions were Eq. (10) for $p = \|, q = \perp$, Eq. (12) for $i = 1, j = 2$, Eq. (15) for $s = x_1, i = 1, 3$ when $\Psi_i = u_i^{\|}$ and $u_i^{\|\prime}$, and for $s = x_2, i = 2, 4$ when $\Psi_i = u_i^{\perp}$, $u_i^{\perp\prime}$, $M_i$ and $V_i$, Eq. (17) and Eq. (19). The rest equations constituted the boundary value problem under Mode II.

### 3.2.4. Analytical dispersion relations

By solving the two eigenvalue problems under the "easy" modes, the dispersion equations, describing the propagation of Bloch waves with a frequency $\omega$ and a wave vector $k_s$ in the soft lattice, could be yielded. Under Modes I ( $s = x_1$ ) and II ( $s = x_2$ ), it yields:

$$P_s(\omega)Q_t(k_t,\omega) + R_s(k_s,\omega)T_t(k_t,\omega) = 0 \tag{20}$$

with $s, t = x_1, x_2$ and $s \neq t$, in which

$$P_s(\omega) = \sin\left(\frac{2l_s\sqrt{\tilde{\rho}_s}}{\sqrt{\tilde{E}_s}}\omega\right),$$

$$Q_t(k_t,\omega) = \gamma_t\alpha_t\beta_t h_t \left(\cos(2k_t l_t) - \cosh(\sqrt{2}l_t\alpha_t)\right)\left(\cos(2k_t l_t) - \cosh(\sqrt{2}l_t\beta_t)\right),$$

$$R_s(k_s,\omega) = h_s\sqrt{2\tilde{E}_s\tilde{\rho}_s}\,\omega\left(\cos(2k_s l_s) - \cos\left(\frac{2l_s\sqrt{\tilde{\rho}_s}}{\sqrt{\tilde{E}_s}}\omega\right)\right), \quad (21)$$

$$T_t(k_t,\omega) = \sinh(\sqrt{2}l_t\beta_t)\alpha_t\left(\cos(2k_t l_t) - \cosh(\sqrt{2}l_t\alpha_t)\right)$$
$$- \sinh(\sqrt{2}l_t\alpha_t)\beta_t\left(\cos(2k_t l_t) - \cosh(\sqrt{2}l_t\beta_t)\right),$$

and $\alpha_t, \beta_t, \gamma_t$ were defined as:

$$\alpha_t = \sqrt{\frac{\tilde{G}_t - B_t\tilde{\rho}_t\omega^2 + \gamma_t}{B_t\tilde{E}_t}}, \quad \beta_t = \sqrt{\frac{\tilde{G}_t - B_t\tilde{\rho}_t\omega^2 - \gamma_t}{B_t\tilde{E}_t}},$$

$$\gamma_t = \sqrt{4B_t\tilde{E}_t\tilde{\rho}_t\omega^2 + (\tilde{G}_t - B_t\tilde{\rho}_t\omega^2)^2}. \quad (22)$$

Specific to non-zero $R_s$ and $Q_t$, Eq. (20) was rewritten as:

$$\frac{P_s(\omega)}{R_s(k_s,\omega)} + \frac{T_t(k_t,\omega)}{Q_t(k_t,\omega)} = 0. \quad (23)$$

The two terms on the left side of the equation denote the contribution of the horizontal and vertical beam components, thereby revealing that in an ideal situation and for any frequency $\omega$, the stretch in the $s$-direction only changed the terms or parameters with subscript $s$, with the remaining unaffected.

For $Q_t = 0$ or $R_s = 0$, the flat bands, i.e., resonance modes independent of the wave vector, were present in different edges of IBZ. For instance, the flat bands under Mode I were located at the frequencies

$$\Gamma - X : f_n = \frac{n}{2l_{x_2}} \sqrt{\frac{\tilde{G}_{x_2} l_{x_2}^2 + B_{x_2} \tilde{E}_{x_2} n^2 \pi^2}{l_{x_2}^2 \tilde{\rho}_{x_2} + B_{x_2} \tilde{\rho}_{x_2} n^2 \pi^2}}, \quad (n = 1, 2...),$$

$$X - M : f_n = \frac{(2n-1)}{4l_{x_1}} \sqrt{\frac{\tilde{E}_{x_1}}{\rho_{x_1}}}, \quad (n = 1, 2...),$$

$$M - Y : f_n = \frac{(2n-1)}{4l_{x_2}} \sqrt{\frac{4\tilde{G}_{x_2} l_{x_2}^2 + B_{x_2} \tilde{E}_{x_2} (2n-1)^2 \pi^2}{4l_{x_2}^2 \tilde{\rho}_{x_2} + B_{x_2} \tilde{\rho}_{x_2} (2n-1)^2 \pi^2}}, \quad (n = 1, 2...),$$

$$Y - \Gamma : f_n = \frac{n}{2l_{x_1}} \sqrt{\frac{\tilde{E}_{x_1}}{\rho_{x_1}}}, \quad (n = 1, 2...). \tag{24}$$

With the increase in the pre-stretch, the geometry and (effective) material of the soft lattice varied significantly. As reported in the existing study on a single soft beam (Zhao and Chang, 2021), a sufficiently large stretch might lead to $B \to 0$ and $\tilde{G} \to \tilde{E}$.

When $B \to 0$, the rotary motion of beam elements was negligible, and Eq. (8) degenerated to the identical form as Eq. (7), i.e.,

$$\tilde{G} u_i^{\perp \prime\prime}(s) + \tilde{\rho} \omega^2 u_i^{\perp}(s) = 0. \tag{25}$$

Thus, the dispersion relation was simplified to a certain extent. For instance, when the beams in both directions were under large stretches, i.e., $B_t \to 0$, the band structure (Eq. (20)) was simplified as:

$$\sqrt{\tilde{G}_t \tilde{\rho}_t} \sin\left(\frac{2l_s \sqrt{\tilde{\rho}_s}}{\sqrt{\tilde{E}_s}} \omega\right) \left(\cos(2k_t l_t) - \cos\left(\frac{2l_t \sqrt{\tilde{\rho}_t}}{\sqrt{\tilde{G}_t}} \omega\right)\right) +$$

$$\sqrt{\tilde{E}_s \tilde{\rho}_s} \sin\left(\frac{2l_t \sqrt{\tilde{\rho}_t}}{\sqrt{\tilde{G}_t}} \omega\right) \left(\cos(2k_s l_s) - \cos\left(\frac{2l_s \sqrt{\tilde{\rho}_s}}{\sqrt{\tilde{E}_s}} \omega\right)\right) = 0. \tag{26}$$

Accordingly, $\tilde{G} \to \tilde{E}$ further simplified the dispersion relation. When $\tilde{G}_s = \tilde{E}_s$, Eq. (26) presented an identical form for $s = x_1$ and $x_2$, which indicated an overlap of the band curves under Modes I and II.

## 4. Band structure of the soft lattice

In the subsequent calculations, the geometry parameters of the unit cell at initial configuration were taken as $2l_{x_1} = 2l_{x_2} = 1\,\text{m}$ and $h_{x_1} = h_{x_2} = 0.1\,\text{m}$. A hyperelastic material with compressible Neo-Hookean strain energy function, i.e., (Chen et al., 2017a)

$$W = \frac{\mathcal{L}}{2}(J-1)^2 - \mu \ln J + \frac{\mu}{2}\left(\text{tr}(F_{si}F_{sj}) - 2\right), \tag{27}$$

was considered, where $\mathcal{L}$ and $\mu$ denote the first and second Lamé constants, respectively. The initial material parameters were assumed with the first and second Lamé constants $\mathcal{L} = 4.32\,\text{MPa}$, $\mu = 1.08\,\text{MPa}$, respectively, as well as the initial mass density $\rho = 1000\,\text{kg/m}^3$. The mentioned parameters denote a compressible variant of material PSM-4 (Bertoldi and Boyce, 2008).

### *4.1 Band structure for typical finite deformation*

The band structure of the soft lattice under the typical finite deformation conditions is illustrated in Fig. 2. To classify the wave modes of each branch, the amount of polarization was defined as (Achaoui et al., 2010):

$$p^2 = \frac{\int |u_{x_1}|^2 dr}{\int \left(|u_{x_1}|^2 + |u_{x_2}|^2\right) dr}, \tag{28}$$

where the integral taken over the current configuration of the unit cell (the cyan-blue area in Fig. 1(c)), $u_{x_1}$ and $u_{x_2}$ represent the displacement components (Fig. 1(c)). For the in-plane wave motion, $p^2 = 0$ or $p^2 = 1$ represent two "easy" wave modes, whereas the value between $0$ and $1$ represents the hybrid mode (In Fig. 2, we display $p^2 \in [0.15,\ 0.85]$ in green without any distinction, which is easy to observe). Then, the "easy" modes predicted with the theoretical model were also provided.

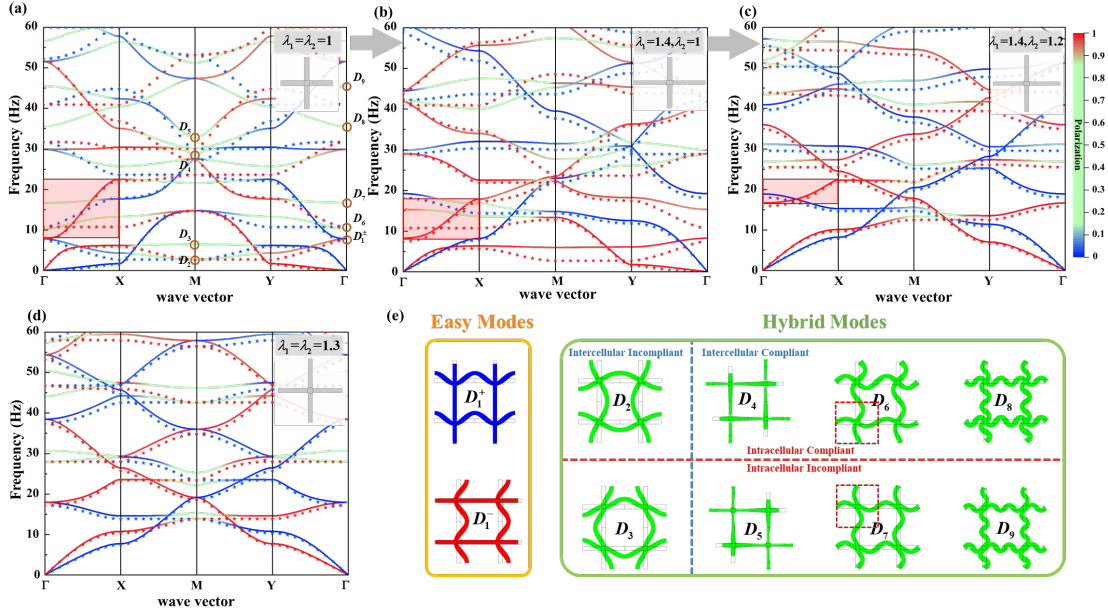

Fig. 2. Phononic band structures for the soft lattice with the typical finite deformation conditions: (a) deformation free ($\lambda_1 = \lambda_2 = 1$), (b) uniaxial stretch ($\lambda_1 = 1.4, \lambda_2 = 1$), (c) biaxial stretch ($\lambda_1 = 1.4, \lambda_2 = 1.2$), and (d) equi-biaxial stretch ($\lambda_1 = \lambda_2 = 1.3$). The solid and scattered lines denote the numerical (FE) and theoretical results, respectively. The typical eigenmodes at high-symmetry points of (a) are illustrated in (e). The colors in the lines and the wave modes indicate the magnitude of the polarization, from 0 (Mode II) to 1 (Mode I). A typical branch in $\Gamma - X$ interval was marked with red shading in (a), (b) and (c), respectively, to show its evolution with the deformation.

The band structure of the deformation-free lattice ($\lambda_1 = \lambda_2 = 1$) was demonstrated in Fig. 2(a). The dispersion curves exhibited the symmetries along the edges ($\Gamma - X - M - Y - \Gamma$) of the IBZ attributed to the geometric symmetry of the unit cell. As indicated from the comparison of the numerical and theoretical results, the theoretical model could reasonably capture the essential characteristics of the propagation of Bloch waves in the soft lattice, especially when the eigenmodes were "easy" ones. However, when the lattice wave propagated under the hybrid mode, the results of the theoretical prediction deviated from the accurate ones.

In Fig. 2(e), several numerically calculated eigenmodes are presented at high-symmetry points $\Gamma$ and $M$. To illustrate the compliance between the unit cells, a supercell containing $2\times 2$ unit cells was utilized for the visualization (with the representative cells framed in red). The two degenerated modes at $D_1$ (Fig. 1(a)) were distinguished by $D_1^+$ and $D_1^-$, which exhibited high compliance and complied with the theoretical prediction of this study. In this case, the eigenfrequencies were well consistent between the numerical and theoretical results. At $D_2 \sim D_9$, the hybrid modes could be distinguished by their intracellular or intercellular compliances. The modes at $D_2$, $D_4$, $D_6$, and $D_8$ were classified as the intracellular compliant modes from the model shapes, whereas those at $D_3$, $D_5$, $D_7$, and $D_9$ were the intracellular incompliant ones. Moreover, the modes at $D_2$ and $D_3$ were demonstrated to be intercellular incompliant, while the others were intercellular compliant. Notably, the intracellular compliant modes were more easily stimulated (or stimulated with a smaller deformation energy) than the incompliant ones, thereby representing lower eigenfrequencies. Correspondingly, the mentioned hybrid modes were prohibited by the "tight binding" model, since they were artificially "degenerated" by constraining the mode shape. The eigenfrequencies of the mentioned "degeneracy" points were generally equated with those of the numerically calculated wave modes with intracellular compliance. Two exceptions could be observed at $D_4$ and $D_8$, and their eigenmodes acquired both intracellular and intercellular compliances, which could be stimulated at lower eigenfrequencies than the theoretical prediction.

The band structures of the lattice under the uniaxial ($\lambda_1 = 1.4$, $\lambda_2 = 1$), biaxial ($\lambda_1 = 1.4$, $\lambda_2 = 1.2$), and equi-biaxial ($\lambda_1 = \lambda_2 = 1.3$) stretches are respectively illustrated in Fig. 2(b)-(d), which are significantly inconsistent with that in Fig. 2(a). Thus, the finite deformation directly impacted the symmetry of the band structure, and the elongation in different tessellation directions diversely impacted the movement of the bands. As indicated from the comparison of Fig. 2(a)-(c), the proportion of the hybrid modes (green curves) in the band structure decreased significantly by the uniaxial or biaxial stretches, which suggested that the finite deformation could "purify" the wave modes. Moreover, the theoretical model here could be more suitable for predicting the band structure of the pre-deformed lattice than that of the undeformed one. For the equi-biaxial stretch (Fig. 2(d)), the geometry and load symmetries could ensure the symmetry of the band structure. Accordingly, all the branches shifted to higher frequencies than that for deformation-free lattice, and the hybrid modes were localized around the high-symmetry points.

Fig. 2(a)-(d) present no complete bandgaps in the focused frequency range. However, the directional bandgaps existed under both wave modes, and their locations moved with the applied finite deformation. Besides, it was noted that the flat band and the degeneracy points were predicted in the theoretical model, whereas they were not present in the numerical results. In the following subsections, all the mentioned phenomena were discussed more specifically.

### *4.2. Effect of pre-deformation on the band structure*
#### *4.2.1. The effect of uniaxial stretch*

To reveal the effect of finite deformation on the soft lattice's band structure, the band diagram was drawn in $\Gamma - X$ interval with uniaxial stretch $\lambda = 1 \sim 1.8$ applied in $x_1$- and $x_2$-directions in Fig. 3(a) and (b), respectively. As revealed from the figure, the stretch along the $x_1$-direction decreased the eigenfrequencies under Mode I, whereas it increases the eigenfrequencies under Mode II. However, the $x_2$-counterpart exerted an opposite effect. Under the hybrid mode, no significant regularity could be observed.

The theoretical model qualitatively explained the evolution pattern. When $s = x_1$, Eq. (26) concluded that $\omega \propto \sqrt{\tilde{E}_{x_1} / \tilde{\rho}_{x_1}} / l_{x_1}$ under Mode I and $\omega \propto \sqrt{\tilde{G}_{x_1} / \tilde{\rho}_{x_1}} / l_{x_1}$ under Mode II. As indicated from the further analysis (Supplementary Material), both $\tilde{G}_{x_1}$ and $l_{x_1}$ increased after a certain tensile deformation, while $\tilde{E}_{x_1}$ and $\tilde{\rho}_{x_1}$ remained almost unchanged. Moreover, the variation of $\sqrt{\tilde{G}_{x_1}}$ was more significant than that of $l_{x_1}$. As a result, the eigenfrequencies under Mode I decreased by the increment of $l_{x_1}$, while that under Mode II increased by the increment of $\sqrt{\tilde{G}_{x_1}}/l_{x_1}$. Likewise, after a stretch along the $x_2$-direction, the eigenfrequencies under Mode I ($\omega \propto \sqrt{\tilde{G}_{x_2}/\tilde{\rho}_{x_2}}/l_{x_2}$) increased by the increment of $\sqrt{\tilde{G}_{x_2}}/l_{x_2}$. In contrast, those under Mode II ($\omega \propto \sqrt{\tilde{E}_{x_2}/\tilde{\rho}_{x_2}}/l_{x_2}$) decreased by the increment of $l_{x_2}$.

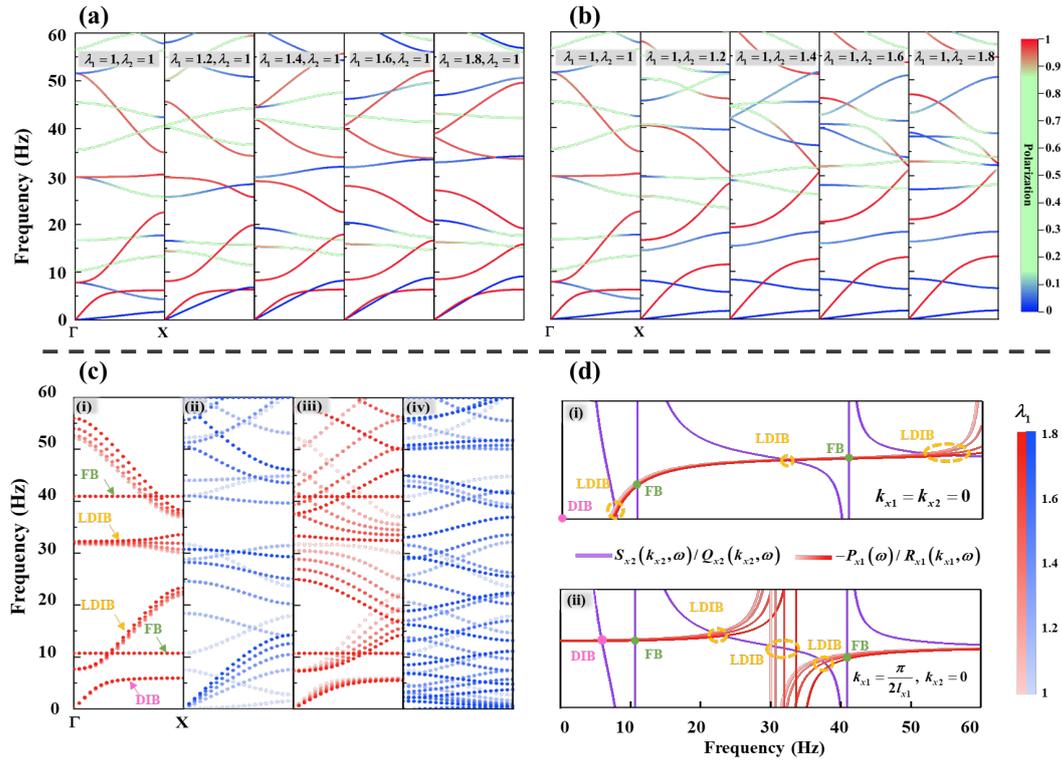

Fig. 3. Effect of uniaxial stretch on the band diagram in the $\Gamma - X$ direction. (a) and (b) represent the numerical results for the deformation applied in the $x_1$- and $x_2$-directions. The colors in the lines indicate the magnitudes of the polarization. (c) illustrates the effect of material (i & ii) and geometrical (iii & iv) factors on the band structure, in which the diagrams for P- (red) and S- (blue) modes are presented independently, and the colors represent the magnitude of the uniaxial stretch. (d) explains the occurrence of the deformation immune bands, the local deformation immune bands, and the flat bands indicated in (c).

The effect of the finite pre-deformation on soft lattice material involved the variations in (effective) material and geometry. Thus, the effects of the mentioned two aspects should be consulted, separately. For this end, under the uniaxial stretch (Fig. 3(a)), the band diagram was recalculated by using the theoretical model with the initial geometric parameters and the effective material parameters after the deformation (Fig. 3(c)). Furthermore, the band diagram obtained by using the current geometric parameters and initial material parameters is presented as a comparison.

The diagram clearly illustrates the distinct effects of geometry and material on the two "easy" modes under the stretch. Moreover, the evolution of the branches could be qualitatively explained by the "tight binding" model. According to Fig. 3 (c-i and ii), the material factor exerted the opposite effects on the two wave modes. Under Mode I, the non-flat bands slightly shifted to higher frequencies with the increment of $\sqrt{\tilde{E}_{x_1}/\tilde{\rho}_{x_1}}$; under Mode II, however, all the eigenfrequencies, primarily affected by $\sqrt{\tilde{G}_{x_1}/\tilde{\rho}_{x_1}}$, increased significantly. However, the effect of the geometric factors was nearly identical for the two modes. In Fig. 3 (c-iii and iv), the reduction in length $l_{x_1}$ decreased all the eigenfrequencies, and the variation was more significant for high-frequency branches than that of the lower ones. By interpreting the numerical result (Fig. 3 (a)) with the mentioned mechanisms, it was concluded that the band structure evolution resulted from the competition between material and geometric factors.

As indicated in Fig. 3 (c-i and iii), several anomalous bands could be perceived in the theoretical results under Mode I, including the deformation immune bands (DIBs), the local deformation immune bands (LDIBs), the flat bands (FBs). As their names suggest, the position of the DIB almost did not vary with the uniaxial stretch in the $x_1$-direction, while in the LDIB, only the eigenfrequency at $\Gamma$ was stationary. In contrast, the dispersionless flat bands were independent of the deformation and the wave number. As indicated from the comparison of Fig. 3 (a) and (c), the theoretically predicted DIBs and LDIBs were consistent with the numerical simulations. However, the flat bands were present in the numerical results. As a replacement, two hybrid modes were generated by hybridizing the flat bands under Mode I and the corresponding non-flat bands under Mode II.

Since the mentioned anomalous bands may have potential applications in bandgap modulation and non-diffractive transmission, the reasons for their occurrence should be analyzed. To this end, according to Fig. 3 (c-i), the function plots of the left two terms of Eq. (23), i.e., $-P_s(\omega)/R_s(k_s,\omega)$ (red curves in Fig. 3 (d)) and $T_t(k_t,\omega)/Q_t(k_t,\omega)$ (purple curves) were drawn, for $s=x_1$, $t=x_2$, $k_{x_2}=0$ and $k_{x_1}=0$ (Fig. 3 (d-i), corresponding to the $\Gamma$ point in Fig. 3 (c-i)) or $\pi/2l_{x_1}$ (Fig. 3 (d-ii), corresponding to the $X$ point). The intersections of the two curves were the solutions of the equation, i.e., the eigenfrequencies of the corresponding bands. It was suggested that $T_t(k_t,\omega)/Q_t(k_t,\omega)$ varied with neither $\lambda_1$ nor $k_{x_1}$. In contrast, $-P_s(\omega)/R_s(k_s,\omega)$ exhibited different change patterns on $\lambda_1$ for various $k_{x_1}$. To be specific, when $k_{x_1}=0$, the uniaxial stretch slightly impacted the function value in a wide frequency range ($0-40\,\text{Hz}$), which achieved the approximately stationary frequencies for the anomalous bands. However, for $k_{x_1}=\pi/2l_{x_1}$, a singularity of $-P_s(\omega)/R_s(k_s,\omega)$ "invaded" this frequency range, so only the solution with lowest frequency retained the immunity to the deformation, which distinguished between DIB and LDIB. Besides, Fig. 3 (c) also clearly demonstrates that FBs (Eq. (24)), attributed to the singularities of $T_t(k_t,\omega)/Q_t(k_t,\omega)$, explaining their immunity to both deformation and wave vector.

*4.2.2 The effect of biaxial stretch*

As indicated from the comparison of Fig. 2(a) and (c), the effect of the biaxial stretch might not be as easy to describe as that of the uniaxial stretch. However, as impacted by the uncoupling of beam components in the orthogonal directions, the effect of a biaxial stretch was equivalent to that of two sequential uniaxial stretches, so it could be qualitatively predicted. For instance, the movement of a representative branch was tracked under Mode I, which was marked with red shading in Fig. (2). A horizontal stretch to $\lambda_1 = 1.4$ made the branch move monotonically downward, except that its left end was almost stationary (as explained in Section 4.2.1), in contrast to Fig. 2(a) and (b). Subsequently, a vertical stretch to $\lambda_2 = 1.2$ made the entire branch shift to higher frequencies (Fig. 2(b) and (c)).

The band diagrams of the soft lattice in $\Gamma - X$ interval for different equi-biaxial states ($\lambda_1 = 1.2 \sim 1.8$) are illustrated in Fig. 3. For comparisons, the band curves predicted by using Eq. (20) and (26) are plotted, thereby indicating that the band curves moved closer to each other with the increase in the magnitude of stretch. They almost coincided with each other when $\lambda_1 = \lambda_2$ was $1.4$. It was therefore indicated that the finite stretch weakened the rotary motion in a soft beam component. Accordingly, as a waveguide, the pre-stretched beam behaved more like a string than a Rayleigh beam. Furthermore, Fig. 3 demonstrates the propinquity of the two easy modes attributed to the agreement of the effective material parameters under the huge stretch, as mentioned at the end of Section 3.2.4 (Eq. (26)). Obviously, the phenomenon could contribute to the modulation of the bandgap to a broad frequency range.

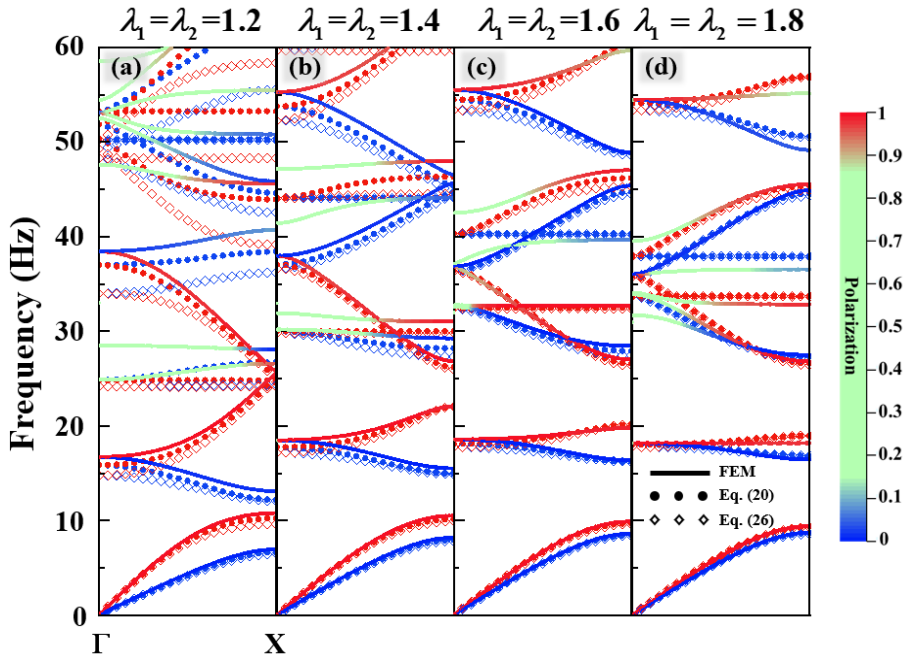

Fig.4. The comparison of the band diagram in the $\Gamma - X$ interval under equi-biaxial stretch,

obtained from numeral simulation (solid lines), Eq. (20) (solid scatters), and Eq. (26) (hollow scatters), respectively.

### *4.3. The evolution of bandgaps under stretch*

The movement of the band curves directly impacted the evolution of the bandgaps. To demonstrate the effect of pre-stretch on the bandgaps, the evolution diagrams of the bandgaps in $\Gamma - X$ interval were drawn under Modes I and II at different stretching states (Fig. 5(a) and (b)). In both figures, the frequency ranges occupied by the hybrid modes were passbands.

Under Mode I (Fig. 5(a)), the uniaxial stretch in the $x_1$-direction narrowed several (the third and fourth) of the bandgaps without affecting their center frequencies, while the $x_2$-counterpart simultaneously increased the bandwidths and center frequencies. As opposed to Mode I, Mode II (Fig. 5(a)) had more bandgaps in this interval. Moreover, the uniaxial stretch in either direction significantly increased the bandwidth, and the center frequencies slightly or significantly increased for the stretches in $x_1$- and $x_2$-directions, respectively. The boundaries of different bandgaps were significantly close under a huge stretch, thereby forming an ultra-wide bandgap with several narrow passbands.

Besides, the broadest in-plane elastic wave bandgap (the bandgap for both modes) appeared in the case of the equi-biaxial stretch (Fig. 5(c)). As indicated from the figure, the bandgaps under Mode I widened dramatically with the increase in the stretch. The two lowest bandgaps for both modes were consistent, thereby forming the widest elastic wave bandgaps at a high strain level. The mentioned phenomenon can be explained as the closer moving of the dispersion curves of the two modes under a large stretch (Eq. (26) and Fig. 4(d)).

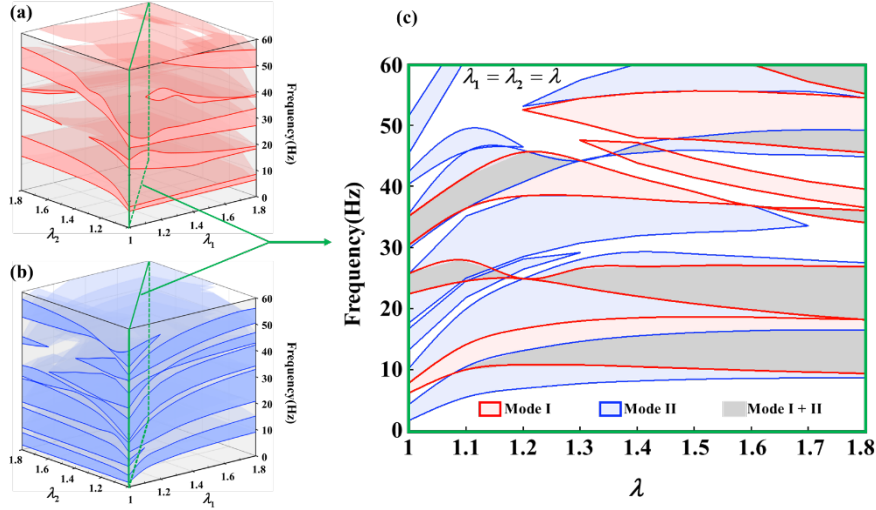

Fig.5. Evolution of bandgaps in $\Gamma - X$ edge of IBZ with different finite deformation. (a) and (b) are the bandgap evolution under Modes I (red) and II (blue). (c) is the same as (a) and (b) but under equi-biaxial stretch. The gray areas in (c) correspond to the elastic wave band-gaps, which is the intersection of the red and blue ones.

## 5. Conclusion and Discussion

In brief, this study investigated the propagation of Bloch waves in a pre-stretched square lattice material. On that basis, the bandgap evolution was determined. As indicated from the numerical simulations, in the lattice material, in-plane elastic waves primarily propagated under two easy (or most compliant) modes, accompanied with some hybrid modes. As the uniaxial pre-stretch increased, the two easy modes demonstrated the monotonous but opposite movements, while the hybrid modes were converted to easy ones and localized to the high-symmetry points. Furthermore, the effect of biaxial stretch could fall to that of two uniaxial stretches in the tessellation directions.

The mechanism underlying the mentioned phenomena was revealed by an elastodynamic "tight binding" model. In this study, a complex system fell to two simpler systems in the model by constraining two easy modes to determine simple dispersion relations exploited for qualitative analyses. Such a degenerated model was suggested to reasonably capture the essential characteristics of the propagation of Bloch waves in easy modes. Moreover, it is conceivable that the model could not accurately predict the hybrid modes. A theoretical analysis was conducted to indicate that the variation of the band structure with the pre-stretch was affected by geometric and material factors, and the decisive parameters of the mentioned two aspects represented the beam component's length and the effective shear modulus, respectively. As impacted by the simple structural configuration of the soft lattice, the uniaxial stretch did not affect the beams in the other direction (except for the joint region), which explained the simple monotonous movement pattern of the simple modes. At a high stretch level, the rotary motion of the beam component under the lateral vibration was weakened. Thus, the wave equation degenerated from the Rayleigh equation to the equation of the string vibration, thereby "purifying" the two modes. Furthermore, in an equi-biaxial stretch, the two easy modes exhibited an approximately coincident band structure attributed to the approximately equal effective Young's modulus and the effective shear modulus.

Besides, the bandgap evolution attributed to the variation in the band structure was demonstrated more specifically, which indicated that the stretches in different directions differently impacted the positions and widths of the bandgaps. The broadest directional elastic wave bandgaps appeared for the equi-biaxial stretch. Though it did not exist in the soft lattice mentioned in this study, the complete bandgap could be implemented by introducing resonance (Liu et al., 2012; Muhammad et al., 2020) or Bragg scattering (Zhang et al., 2020) mechanisms.

In addition, the comparing numerical and theoretical calculations indicated the differences between the physical and simplified theoretical models. It was therefore demonstrated that the theoretically predicted flat bands, the characteristics showing promising applications in topological acoustics (Ge et al., 2018), were not presented in the numerical calculations. However, it was not excluded that such characteristics could be determined by means (e.g., tuning the slenderness of the beam components (Bordiga et al., 2019)).

Hopefully, this study may present more insights into the elastodynamics of soft lattice materials and into the ways their behavior can be exploited in engineering design. Furthermore, the authors hope that this study can elucidate the design and optimization of lightweight, tunable elastic wave manipulation devices.

## Acknowledgments

This work was supported by the National Natural Science Foundation of China (Grant No.11602294) and the 2115 Talent Development Program of China Agricultural University.